\documentclass[twocolumn]{aa}  

\usepackage{graphicx}
\usepackage{txfonts}
\usepackage{color}
\usepackage{xcolor}
\usepackage{multicol}
\usepackage{amsmath,mathrsfs}
\usepackage{makecell}
\usepackage{subcaption}
\usepackage{pdflscape}
\usepackage[normalem]{ulem}
\usepackage[skip=2pt plus1pt, indent=15pt]{parskip}
\usepackage{threeparttablex}
\usepackage{tikz}
\usepackage[colorlinks=true, linkcolor=black, citecolor=blue, urlcolor=magenta]{hyperref}
\newcommand{\crir}  {\zeta^{\rm ion}_{{\rm H}_2}}

\begin{document}

   \title{Parsec-scale cosmic-ray ionisation rate in Orion}


   \author{A.~Socci
          \inst{1}
          \and
          G.~Sabatini\inst{2}
          \and
          M.~Padovani\inst{2}
          \and
          S.~Bovino\inst{3,4,5}
          \and
          A.~Hacar\inst{1}
          }

   \institute{Institute for Astronomy (IfA), University of Vienna,
              T\"urkenschanzstrasse 17, A-1180 Vienna\\
              \email{andrea.socci@univie.ac.at}
         \and
             INAF, Osservatorio Astrofisico di Arcetri, Largo E. Fermi 5, 50125 Firenze, Italy
          \and
             Chemistry Department, Sapienza University of Rome, P.le A. Moro, 00185 Rome, Italy
         \and
             INAF, Istituto di Radioastronomia, Via Gobetti 101, I-40129 Bologna, Italy
        \and
            Departamento de Astronomía, Facultad Ciencias Físicas y Matemáticas, Universidad de Concepción Av. Esteban Iturra s/n Barrio Universitario, Casilla 160, Concepción, Chile
            }

   \date{---}

 
  \abstract
   {Cosmic rays are a key component of the interstellar medium since they regulate the dynamics and the chemical processes in the densest and coldest regions of molecular clouds. Still, the cosmic-ray ionisation rate of H$_2$ ($\crir$) is one of the most debated parameters characterising molecular clouds, due to the uncertainties in the adopted chemical networks and the analysis techniques.}
   {This work aims to homogeneously estimate the $\crir$ at parsec scales, towards the Orion Molecular Clouds OMC-2 and OMC-3. We explore the change in $\crir$ across a whole star-forming region by probing a range of column densities never explored before. The significant increase in statistics obtained by studying an entire region allows us to put stronger constraints on the range of $\crir$ values and exploit its connection with the physical properties of the ISM.}
   {The most recent $\crir$ estimates are based on o$-$H$_2$D$^+$, a direct product of the interaction between cosmic rays and H$_2$ in cold clouds. Since observations of o$-$H$_2$D$^+$ are challenging, in this work we proxy its abundance through CO depletion by employing C$^{18}$O (2$-$1) observations towards OMC-2 and OMC-3, taking advantage of the existing correlation between the two parameters. Using additional observations of HCO$^+$ (1$-$0) and DCO$^+$ (3$-$2), we determine the deuteration fraction and, we finally derive the map of $\crir$ in these two regions.}
   {The C$^{18}$O depletion correlates with both the total column density of H$_2$ and the N$_2$H$^+$ emission across OMC-2 and OMC-3. The obtained depletion factors and deuteration fractions are consistent with previous values obtained in low- and high-mass star-forming regions. In addition, they show a functional dependence within the coldest fields in our sample. The cosmic-ray ionisation rate we derive shows values of $\crir\sim5\times10^{-18}-10^{-16}$~s$^{-1}$, in good agreement with previous estimates based on o$-$H$_2$D$^+$ observations. The $\crir$ also shows a functional dependence with the column density of H$_2$ across a full order of magnitude ($\sim10^{22}-10^{23}$~cm$^{-2}$). The estimated values of $\crir$ overall decrease for increasing $N(\mathrm{H_2}$), consistently with the predictions from theoretical models.}
   {The approach we follow delivers results comparable with theoretical predictions and previous independent studies, confirming both the robustness of the analytical framework as well as CO depletion being a viable proxy of o$-$H$_2$D$^+$. We also explore the major limitations of the method by varying the physical size of the gas crossed by the cosmic rays (i.e., the path length). By employing a path length obtained from low resolution observations, we recover values of the $\crir$ well below any existing theoretical and observational prediction. This discrepancy suggests interferometric observations as mandatory in order to reliably constrain the $\crir$ also at parsec scales.}


   \maketitle
%
\captionsetup{labelfont=bf}
\renewcommand{\ttdefault}{pcr}

\section{Introduction}

Cosmic rays (CRs) are a fundamental component of the interstellar medium (ISM), yet, possibly, the most puzzling. Low-energy CRs ($E<1$ TeV) can penetrate the cold ($T\lesssim20$ K), dense gas ($n\gtrsim10^4$ cm$^{-3}$) of molecular clouds and interact with H$_2$. The chain reactions following the H$_2$ ionisation produce H$_3^+$ \citep[see][for a review]{indi12}, a key molecule responsible for the production of most ionic species within molecular clouds \citep{herbst73}. The amount of ions over the number of neutral species, namely the ionisation fraction, is of paramount importance for star formation, as it regulates the coupling between magnetic fields and the gas \citep{pado20}, slowing or even halting the gravitational collapse of the cloud \citep{bergin07}. In addition, the amount of ions, H$_3^+$ in particular, drives the chemical evolution of the coldest regions within a cloud \citep[e.g.,][]{caselli12}. However, while observed in infrared absorption in diffuse clouds \citep{indi12,oka19}, H$_3^+$ cannot be detected in the mm regime, preventing a direct measure of the degree of ionisation in dense molecular clouds. The cosmic-ray ionisation rate (CRIR), directly connected to H$_3^+$, has been therefore determined over the years through several proxies, such as radicals \citep{black77}, ions \citep{caselli98,cecca14,reda21}, neutrals \citep{fontani17,favre18}, and a combination of the latter \citep[e.g.,][]{luo23}. The values of CRIR determined in molecular clouds span three orders of magnitude ($\sim10^{-17}-10^{-14}$~s$^{-1}$) and show a dependence with the column density of H$_2$, $N(\mathrm{H_2})$, as expected from theoretical models \citep{pado09,pado22}. Yet, the variety of techniques (most of them model-dependent), tracers and resolutions makes for a non-homogeneous sample of scattered estimates, especially in dense clouds \citep[$N(\mathrm{H_2})\gtrsim10^{22}$~cm$^{-2}$; e.g.,][]{BovinoGrassi24}.

\citet{bovino20} has recently proposed ortho$-$H$_2$D$^+$ (hereafter o$-$H$_2$D$^+$) as a proxy for the CRIR relative to molecular hydrogen ($\crir$). H$_2$D$^+$ is the first product of the deuteration of H$_3^+$ and its \textit{ortho} form has been successfully detected in the millimetre regime at both low \citep{caselli03,caselli08,gianne19,saba20,mietti20,bovino21} and high resolution \citep{reda21b,reda22,saba23}. Since the deuterium enrichment is favoured in cold dense gas \citep{millar89,walmsley04}, o$-$H$_2$D$^+$ is a reliable proxy of the $\crir$ in young pre-stellar environments \citep[$t\lesssim0.2$~Myr;][]{bovino20}. This analytic derivation has been recently employed by \citet{saba20} and \citet{saba23}, who determined $\crir$ from o$-$H$_2$D$^+$ in a sample of high-mass clumps and cores. The ionisation rates obtained by the authors are consistent with the theoretical predictions for a high-density regime \citep{padovani24}. Whilst the estimated $\crir$ reproduces the theoretical expectations, the detection of o$-$H$_2$D$^+$ has been a challenge for more than a decade \citep[e.g.,][]{caselli08}. As a consequence, the method proposed by \citet{bovino20} has been applied only to a limited range of column densities and scales, in particular for $N(\mathrm{H_2})\gtrsim10^{23}$~cm$^{-2}$ and $R\lesssim0.5$~pc, respectively \citep{saba23}. 

The present study aims to derive the column density of o$-$H$_2$D$^+$ and then the $\crir$ for column densities within $N(\mathrm{H_2})\sim10^{22}-10^{23}$~cm$^{-2}$ and for volume densities around $n\sim10^5$~cm$^{-3}$. In the cold gas of molecular clouds where these volume densities are achieved, CO depletes onto dust grains, ions become abundant and deuteration processes are overall boosted \citep{caselli12}. In this context, the abundance of H$_2$D$^+$ is expected to grow for increasing depletion factors, as also observed in different samples of both low- \citep{caselli08} and high-mass \citep{saba20} star-forming regions. Following the aforementioned studies, in this work we determine $N$(o$-$H$_2$D$^+$) from the C$^{18}$O depletion factor to provide an independent estimate of $\crir$ on parsec scales. Despite its limitations, discussed throughout the paper, our approach produces results consistent with previous works, while extending the dynamic range of column densities explored so far in the $\crir$ estimation.

The paper is then structured as follows: we introduce the method for the $\crir$ determination and its main assumptions (Sect.~\ref{sec:method}); then, we present the observations, both novel (Sect.~\ref{sec:observations}) and archival (Sect.~\ref{subsec:sourcesel}), used throughout the paper and the related source selection; next, we describe the analysis and results obtained for the degree of depletion (Sect.~\ref{subsec:depletion}), the expected abundance of o$-$H$_2$D$^+$ (Sect.~\ref{subsec:h2dabund}), the deuteration fraction (Sect.~\ref{subsec:deutfrac}), and the ionisation rate (Sect.~\ref{subsec:zeta}), along with their error budget (Sect.~\ref{subsec:errorbud}) and limits of applicability (Sect.~\ref{subsec:desrate}); finally, we discuss the connection between $\crir$ and $\mathrm{N(H_2)}$ both from our analysis and in comparison with the theoretical predictions, together with the major limitations to the ionisation rate estimate (Sect.~\ref{sec:discussion}). Section~\ref{sec:conclusions} wraps up the major findings of the present study.

\section{Method}\label{sec:method}

The method introduced by \citet{bovino20} provides a model-independent analytical expression to estimate $\crir$, which reads as follows \citep[see also][]{saba23}:
\begin{equation}
    \crir = k_{\rm{CO}}^{\rm{o-H_3^+}}\frac{N({\rm{o-H_2D^+}}\times N({\rm{CO}})}{3~R_{\rm{D}}\times N({\rm{H_2}})\times l}.
    \label{eq:saba_eq}
\end{equation}
In the above equation: $k_{\rm{CO}}^{\rm{o-H_3^+}}$ is the destruction rate of o$-$H$_3^+$ by CO, considered here as main destruction path for o$-$H$_3^+$ \citep{reda24}; $N({\rm{CO}})$ is the total column density of CO; $R_{\rm{D}}$ ($=$~$N(\mathrm{DCO^+})/N(\mathrm{HCO^+})$) is the deuteration fraction; $l$ is the path length over which CO suffers depletion and we estimate the column densities of our tracers; $N({\rm{o-H_2D^+}})$ and $N({\rm{H_2}})$ are the column densities of o$-$H$_2$D$^+$ and H$_2$, respectively. 

The robustness and reliability of Eq.~(\ref{eq:saba_eq}) have been recently confirmed by \citet{reda24} through synthetic observations of the molecular tracers simulated in a three-dimensional grid. The analysis provided $\crir$ estimates accurate within a factor of 2-3 compared to the input value of the simulation. These findings confirm the confidence level suggested by \citet{bovino20} for the method. 

To apply Eq.~(\ref{eq:saba_eq}), our study relies on two main assumptions. First, we assume o$-$H$_2$D$^+$ to be efficiently traced by the C$^{18}$O depletion factor ($f_\mathrm{D}$). This hypothesis is based on the correlation between the o$-$H$_2$D$^+$ abundance and the degree of CO depletion, a result consistently derived in earlier studies both through observations \citep[e.g.,][]{crapsi05,caselli08,saba20} and through theoretical predictions \citep[e.g.,][]{sipi15,bovino19}. We will therefore proxy the abundance of o$-$H$_2$D$^+$ via $f_\mathrm{D}$ using the correlation reported by \citet{saba20}. 

Second, we make an assumption on the path length $l$ in Eq.~(\ref{eq:saba_eq}). This path length is the physical extent of the o$-$H$_2$D$^+$ emission along the line of sight. Largely unknown, $l$ is usually taken as the physical size of the o$-$H$_2$D$^+$ emission in the plane of the sky \citep[e.g.,][]{saba23} or in the box where it is simulated \citep[e.g.,][]{reda24} and constitutes one of the main limitations of the present method. We make use of $l=0.050\pm0.015$~pc, the typical width of filamentary substructures seen in N$_2$H$^+$ across OMC-2 and OMC-3 (Socci et al. submitted). N$_2$H$^+$ is a molecular ion which becomes abundant when CO freezes-out on grains \citep{bergin07} and whose emission has been proven to correlate with the one of o$-$H$_2$D$^+$ \citep{reda22}. Thus, we take it as the representative scale for our study. Although unresolved in our observations (see Sect.~\ref{sec:observations}), this scale is the most accurate estimate in the regions without relying on direct o$-$H$_2$D$^+$ observations. In Sect.~\ref{sec:discussion}, we will discuss how different assumptions on $l$ affect the determination of $\crir$. 

\section{Observations}

\subsection{IRAM-30m observations}\label{sec:observations}

\begin{figure*}
    \centering
    \includegraphics[width=0.9\linewidth]{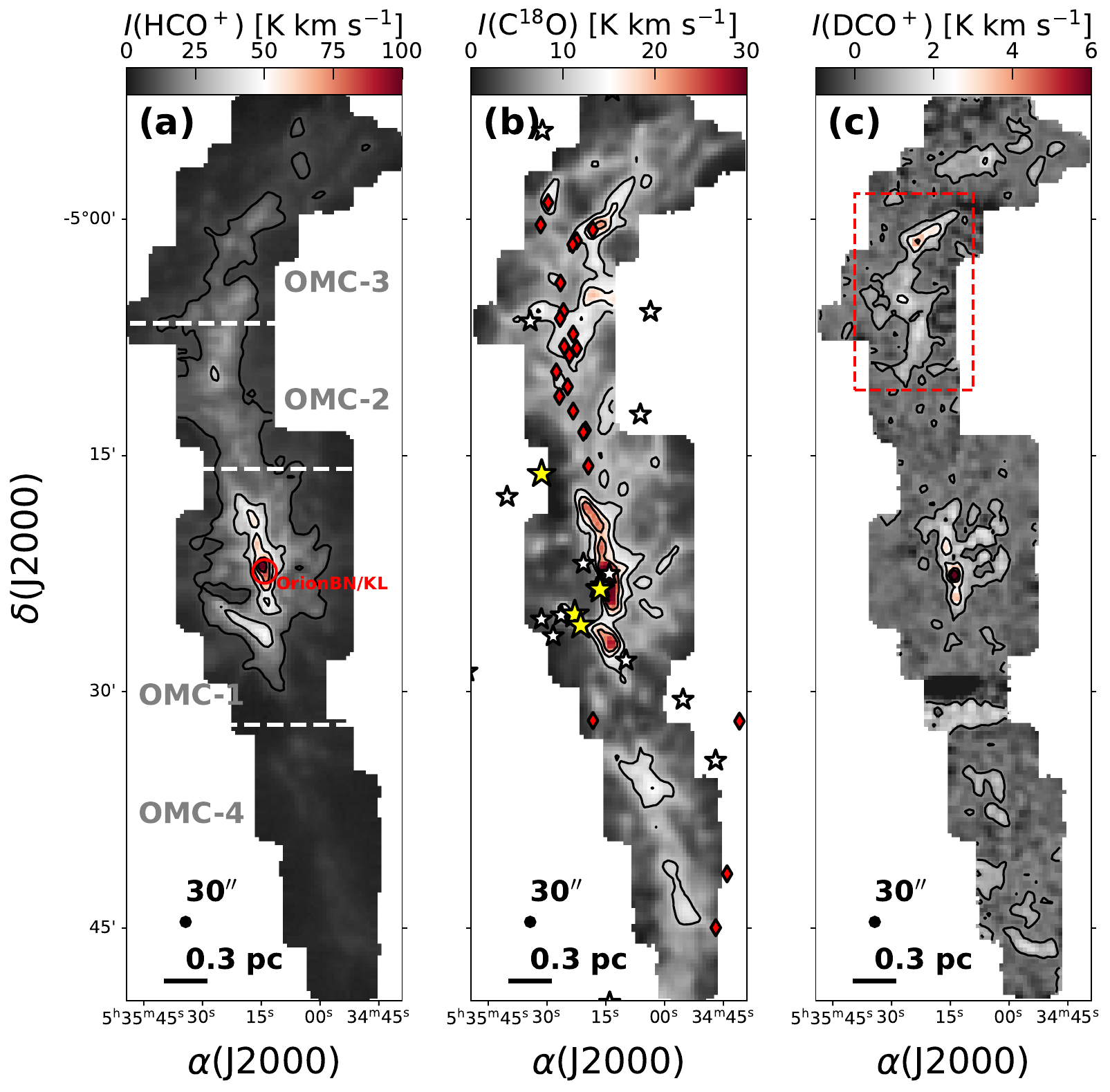}
    \caption{The Integral Shaped Filament as seen through our suite of tracers. \textbf{Panel (a)}: HCO$^+$ (1$-$0) integrated intensity map of the four Orion Molecular Clouds (OMCs) composing the ISF \citep[separated by white lines;][]{bally08}. The contours displayed correspond to [10, 30, 50, 70, 90]~K~km~s$^{-1}$. The red cicle represents the position of the Orion BN/KL object. \textbf{Panel (b)}: C$^{18}$O (2$-$1) integrated intensity map of the OMCs composing the ISF. We include here the O-type (yellow stars) and B-type (white stars) stars in the region, gathered from the SIMBAD Astronomical Database \citep{wenger00}, and the outflows detected by \citet{tanabe19} (red diamonds; see text for full discussion). The contours displayed correspond to [10, 15, 20]~K~km~s$^{-1}$. \textbf{Panel (c)}: DCO$^+$ (3$-$2) integrated intensity map of the OMCs composing the ISF. The contours displayed correspond to [0.5, 2, 4, 6]~K~km~s$^{-1}$, where the lowest corresponds to a signal-to-noise ratio of 3. The red box is the region selected for the analysis, comprising part of OMC-2 and the whole OMC-3 (see text for a discussion).}
    \label{fig:raw_data}
\end{figure*}

Applying the method from \citet{bovino20} to estimate $\crir$ outside high-mass cores and clumps implies exploring column densities $N(\mathrm{H_2})\lesssim10^{23}$~cm$^{-2}$ at parsec scales. To fulfil this goal, the Intregral Shaped Filament (ISF) in Orion \citep{bally87} appeared as a natural choice. The ISF is a $\sim13$ pc long filament in the Orion A molecular cloud, prominent in both dust continuum \citep[e.g.,][]{john99} and molecular lines \citep[e.g.,][]{bally87}, and composed by several condensations, the Orion Molecular Clouds \citep[OMCs, see][for a review]{bally08}. The ISF is also the closest site of high-mass star-formation in the Solar neighbourhood \citep[414 pc;][]{menten07} harbouring the Orion Nebula Cluster \citep[ONC;][]{hill98} within OMC-1. The extended size, the large dynamic range in  H$_2$ column densities and the connection to high-mass star formation make the ISF the best candidate for the present study.

We surveyed OMC-1, OMC-2, OMC-3 and OMC-4 in C$^{18}$O (2$-$1) (219.560~GHz), DCO$^+$ (3$-$2) (216.113~GHz) and HCO$^+$ (1$-$0) (89.189~GHz) using single-dish observations at 1mm and 3mm. The observations were carried out at the 30-metre Institut de Radioastronomie Millimétrique telescope (IRAM-30m) in Granada (Spain) during November 2013 (Project: 032-13, PI: A. Hacar). The large-scale IRAM-30m mosaics centred on Orion BN/KL \citep[][$\alpha$(J2000),$\delta$(J2000) = 05h:35m:14s.20, -05$^{\circ}$22$'$21$''$.5; see Fig.~\ref{fig:raw_data}, panel a]{genzel89} are obtained with single $200''\times200''$ or $100''\times100''$ Nyquist-sampled On-The-Fly (OTF) maps in Position Switching (PSw) mode \citep[see][for additional information on the observations]{hacar17,hacar20}. Our suite of tracers have been observed with the same receiver, the Eight Mixer Receiver \citep[EMIR;][]{carter12}, but with different backends: C$^{18}$O (2$-$1) was observed with the VErsatile SPectrometer Array (VESPA) at 20~kHz resolution ($\sim0.027$~km~s$^{-1}$ at 219.560~GHz), while DCO$^+$ (3$-$2) and HCO$^+$ (1$-$0) with the Fast Fourier Transform Spectrometer \citep[FTS;][]{klein12} at 200~kHz resolution ($\sim0.27$~km~s$^{-1}$ at 216.113~GHz and $\sim0.66$~km~s$^{-1}$ at 89.189~GHz, respectively). The spectra obtained in antenna temperature ($T_\mathrm{a}^*$) have been converted in main beam temperature ($T_\mathrm{mb}$) with the relation $T_\mathrm{mb} = T_\mathrm{a}^*~F_\mathrm{ef}/B_\mathrm{ef}$. Here we used the standard forward ($F_\mathrm{ef}$) and backward ($B_\mathrm{ef}$) efficiencies for the IRAM-30m telescope\footnote{\textbf{\url{https://publicwiki.iram.es/Iram30mEfficiencies}}} (C$^{18}$O (2$-$1): $B_\mathrm{ef}$ = 0.61, $F_\mathrm{ef}$ = 0.93; HCO$^+$ (1$-$0): $B_\mathrm{ef}$ = 0.81, $F_\mathrm{ef}$ = 0.95; DCO$^+$ (3$-$2): $B_\mathrm{ef}$ = 0.62, $F_\mathrm{ef}$ = 0.93). Finally, all the observations have been convolved to a final resolution of 30$''$ ($\sim0.06$ pc at the Orion distance). Figure \ref{fig:raw_data} shows the integrated intensity maps of the three molecules obtained by integrating the spectral cubes in the same velocity range $\Delta V_\mathrm{lsr} = [6, 15]$~km~s$^{-1}$, around the typical velocity expected for the ISF \citep[$V_\mathrm{lsr}\sim10$~km~s$^{-1}$;][]{bally08}. Similarly, the rms noise has been estimated across the whole map, integrated in spectral windows with same velocity range but devoid of emission.

C$^{18}$O and HCO$^+$ are readily detected and show widespread emission across the ISF. The high-sensitivity IRAM-30m observations grant a peak-to-noise dynamic range of $\gtrsim100$ for the intensities of both tracers. The more extended component of C$^{18}$O and HCO$^+$ emission across the ISF shows values on average of $\sim6-10$~K~km~s$^{-1}$. These values increase along the ISF crest up to peaks of $\sim35$~K~km~s$^{-1}$, for C$^{18}$O and $\sim100$~K~km~s$^{-1}$, for HCO$^+$, seen towards OMC-1. DCO$^+$, on the other hand, shows a limited peak-to-noise dynamic range of $\sim50$, along with a patchy and uneven emission across the ISF. The few peaks with intensities of $\gtrsim4$~K~km~s$^{-1}$ are localised towards OMC-1 and OMC-2/OMC-3. All three tracers show their peak intensities towards OMC-1, where the radiation coming from the Trapezium stars \citep{hill97} and the ONC heats up the gas to temperatures above $\sim40$~K \citep{friesen17}. While all tracers share OMC-1 as the region of peak intensity, the C$^{18}$O emission is often anti-correlated across the ISF compared to the emission of HCO$^+$. This change in morphology is prominent towards two regions: first, along OMC-2/OMC-3, where the C$^{18}$O emission is scattered and fragmented while HCO$^+$ shows a coherent filamentary shape (see Fig.~\ref{fig:raw_data}, panels a, b); second, in OMC-4, where C$^{18}$O emission is instead brighter compared to the one of HCO$^+$. The behaviour in OMC-2/OMC-3 is of interest as it hints towards the presence of extended C$^{18}$O depletion in these two regions.

\subsection{Archival observations and source selection}\label{subsec:sourcesel}

The ISF is a well-known region with a plethora of archival data available. For our study, we will take advantage of a few of these ancillary observations with comparable resolution to our data. Among these, we are interested in a $N(\mathrm{H_2})$ map of the region to determine the candidate sites for depletion. Large scale depletion is in fact expected for column densities $N(\mathrm{H_2})\gtrsim10^{22}$~cm$^{-2}$ \citep{bergin07,tafalla23}. The column density map of H$_2$ has been obtained from the 850 $\mu$m dust opacity, observed by the \textit{Herschel} and \textit{Planck} observatories, following the prescriptions of \citet{lombardi14} (see also Appendix~\ref{sec:addmaps}). Figure~\ref{fig:aux_maps}, panel (a), shows the ISF as seen in column density of H$_2$ (36.2$''$). The map shows material with column densities $N(\mathrm{H_2})\geq10^{22}$~cm$^{-2}$ extended across the whole filament with peaks of $N(\mathrm{H_2})\sim10^{23}$~cm$^{-2}$ along its crest, in correspondence of the OMCs. We will focus on these regions, where the column density has a dynamic range of almost two orders of magnitude and large CO depletion is expected.

As mentioned in the previous section, N$_2$H$^+$ emission strongly anti-correlates with the one of CO, as it gets destroyed by the latter \citep[e.g.,][]{tobin13}. To further assess the presence of large-scale depletion in the ISF, we qualitatively compare the total column density of H$_2$ with the IRAM-30m observations of N$_2$H$^+$ (1$-$0) from \citet{hacar17} (Fig.~\ref{fig:aux_maps}, panel b). Although not as extended as the material with $N(\mathrm{H_2})\sim10^{22}$~cm$^{-2}$, N$_2$H$^+$ has emission widespread across the ISF and closely correlated with $N(\mathrm{H_2})$ towards the OMCs (panel a). All four regions are thus candidates for efficient depletion and, as a consequence, good targets to estimate $\crir$.

Our focus is instead shifted towards OMC-2 and OMC-3 only, as suggested previously. There are two main reasons to consider only these two among the four OMCs. First, the OMC-1 region hosts several O-B stars \citep[stars in Fig.~\ref{fig:aux_maps}, panel b;][]{wenger00} and a plethora of highly embedded outflows \citep[e.g.,][]{rivilla13}. The radiation and heating from these objects can modify the C$^{18}$O abundance by its selective photodissociation or its release in the gas phase from the ice coatings of dust grains \citep[see][for a more thorough discussion]{draine11}. For a reliable estimate of the depletion, we have to exclude OMC-1 where the above processes are the most efficient in the ISF. Second, DCO$^+$ (3$-$2) has weak emission across the ISF (see Sect.~\ref{sec:observations}) and, in particular, no emission above a signal-to-noise of 3 towards OMC-4. We are therefore limited in our analysis to OMC-2 and OMC-3 \citep[see][for a review]{peterson08}.

OMC-2 and OMC-3 themselves show more than an order of magnitude variation in $N(\mathrm{H_2})$ and temperatures between $T\sim10-40$~K, according to different estimates \citep{friesen17,hacar20}. Since outflows are also detected in these two regions \citep[diamonds in Fig.~\ref{fig:raw_data}, panel b;][]{tanabe19}, we minimise their contamination to the C$^{18}$O depletion estimate by imposing a temperature cut in the analysis of $T_\mathrm{K} = 10-25$~K \citep[using $T_\mathrm{K}(\mathrm{NH_3})$ at 32$''$, Fig.~\ref{fig:aux_maps}, panel c;][]{friesen17}. Given the close correlation between NH$_3$ and N$_2$H$^+$ \citep[e.g.,][]{tafalla02}, we expect $T_\mathrm{K}$(NH$_3$) to be the representative temperature for the gas in which CO freezes-out onto the grains. The range $T_\mathrm{K} = 10-25$~K is consistent with those previously explored in studies of $f_\mathrm{D}$ in high-mass regions \citep[e.g.,][]{saba19}. In addition, $\sim20-25$~K corresponds to the temperature limit for which CO evaporates from dust grains \citep{bergin07}, given $n\gtrsim10^5$~cm$^{-3}$, volume densities we expect to explore with the present work.

OMC-2 and OMC-3 match all the previous characteristics with a clear detection of DCO$^+$, a low number of O-B stars nearby, a few scattered outflows, column densities up to $N(\mathrm{H_2})\sim10^{23}$ cm$^{-2}$ and temperatures mostly below $\lesssim25$~K. Across these $\sim1-2$~pc towards the North of the ISF, we expect C$^{18}$O to be depleted and not photodissociated by stellar activity. Under such conditions, the depletion is expected to correlate to the o$-$H$_2$D$^+$ abundance \citep{caselli08, saba20}, and, therefore, it may be used to estimate the $\crir$.

\section{Analysis and Results}\label{sec:analysis}

\subsection{The depletion map of OMC-2 and OMC-3}\label{subsec:depletion}

\begin{figure*}
\centering
\includegraphics[width=\linewidth]{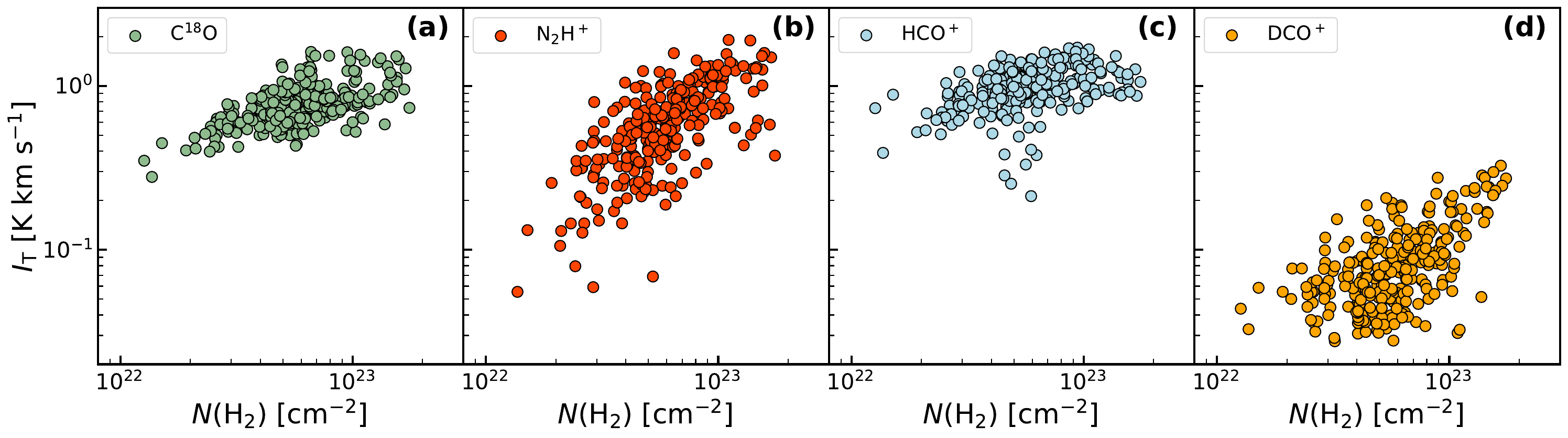}
\caption{Temperature-corrected integrated intensities ($I_\mathrm{T}$) for our suite of tracers: C$^{18}$O (panel a), N$_2$H$^+$ (panel b), HCO$^+$ (panel c) and DCO$^+$ (panel d). All the $I_\mathrm{T}$ values are computed following \citet{tafalla21}, and compared, pixel-by-pixel, to $N(\mathrm{H_2})$ \citep{lombardi14}.}
\label{fig:tracers_comparison}
\end{figure*}

As we want to determine the onset of depletion, properly estimating the column density of C$^{18}$O ($N(\mathrm{C^{18}O})$) is mandatory. First insights in this direction come from the inspection of both $I(\mathrm{C^{18}O})$, $I(\mathrm{N_2H^{+}})$ and their comparison with $N(\mathrm{H_2})$. We can qualitatively explore the $N(\mathrm{H_2})$ regime for which C$^{18}$O depletes by computing the temperature corrected line intensity \citep{tafalla21}:
\begin{equation}\label{eq:tcorrint}
    I_{\rm{T}} = \frac{I}{J_{\nu}(T_{\rm{K}}) - J_{\nu}(T_{\rm{bg}})}.
\end{equation}

Through Eq.~(\ref{eq:tcorrint}), the column density of H$_2$ is now the only factor contributing to the line excitation, therefore to its intensity $I_\mathrm{T}$. Figure \ref{fig:tracers_comparison} shows the comparison between $I_\mathrm{T}$ and $N(\mathrm{H_2})$ for our suite of tracers. We will now focus on C$^{18}$O (panel a) and N$_2$H$^+$ (panel b): these two tracers show a continuous distribution with $N(\mathrm{H_2})$, however they behave differently for increasing column densities. Across the dynamic range probed by our analysis ($N(\mathrm{H_2})\sim10^{22}-10^{23}$~cm$^{-2}$), the flattening of $I_\mathrm{T}(\mathrm{C^{18}O})$, on one hand, and the linear increase of $I_\mathrm{T}(\mathrm{N_2H^{+}})$, on the other, represent a clear chemical differentiation of the two tracers and, in particular, the effect of depletion on C$^{18}$O \citep[e.g.,][]{tafalla23}.

To quantitatively describe the degree of C$^{18}$O depletion in OMC-2 and OMC-3, we need to determine its column density ($N(\mathrm{C^{18}O})$). The latter can be computed as follows \citep{gold99}:
\begin{equation}
    N_{\rm{tot}} = \frac{8\pi k\nu^2}{hc^3A_{\rm{ul}}}\frac{Q(T_{\rm{ex}})~e^{E_{\rm{ul}}/kT_{\rm{ex}}}}{g_{\rm{u}}}\frac{I~\tau}{1-e^{-\tau}},
    \label{eq:coldens}
\end{equation}
with the corresponding parameters: $h$, the Planck constant; $k$, the Boltzmann constant; $c$, the light speed; $\nu$, the rest frequency of the line; $A_\mathrm{ul}$, the Einstein coefficient for spontaneous emission; $Q$, the partition function calculated at the excitation temperature of the line ($T_\mathrm{ex}$); $g_\mathrm{u}$, the degeneracy of the upper level; $E_\mathrm{u}$, the excitation energy of the upper level; $\tau$, the optical depth of the line. We can simplify the above equation for $N(\mathrm{C^{18}O}$) by introducing two approximations. First, we can assume C$^{18}$O (2$-$1) to be in Local Thermodynamic Equilibrium (LTE) at the gas kinetic temperature $T_\mathrm{K}$. Given the presence of N$_2$H$^+$ in the gas phase, the volume densities are expected to be $n\sim10^5$~cm$^{-3}$, which is at least an order of magnitude larger than the critical density of C$^{18}$O \citep[e.g.,][]{tafalla02}. We therefore take $T_\mathrm{ex}(\mathrm{C^{18}O}) = T_\mathrm{K}(\mathrm{NH_3})$ for the rest of the analysis. Second, we work under the optically thin assumption (i.e., $\tau<<1$) adducing the flattening in C$^{18}$O intensity only to its freeze-out onto the dust grains \citep[see][for a discussion]{tafalla23}. With these two approximations, Eq.~(\ref{eq:coldens}) can be rewritten as:
\begin{equation}
    N_{\rm{tot}}({\rm{C^{18}O}}) \sim \frac{8\pi k\nu^2}{hc^3A_{\rm{ul}}}\frac{Q(T_{\rm{K}})~e^{E_{\rm{ul}}/kT_{\rm{K}}}}{g_{\rm{u}}}I({\rm{C^{18}O}}).
    \label{eq:c18odens}
\end{equation}

\begin{figure*}
\centering
\includegraphics[width=\linewidth]{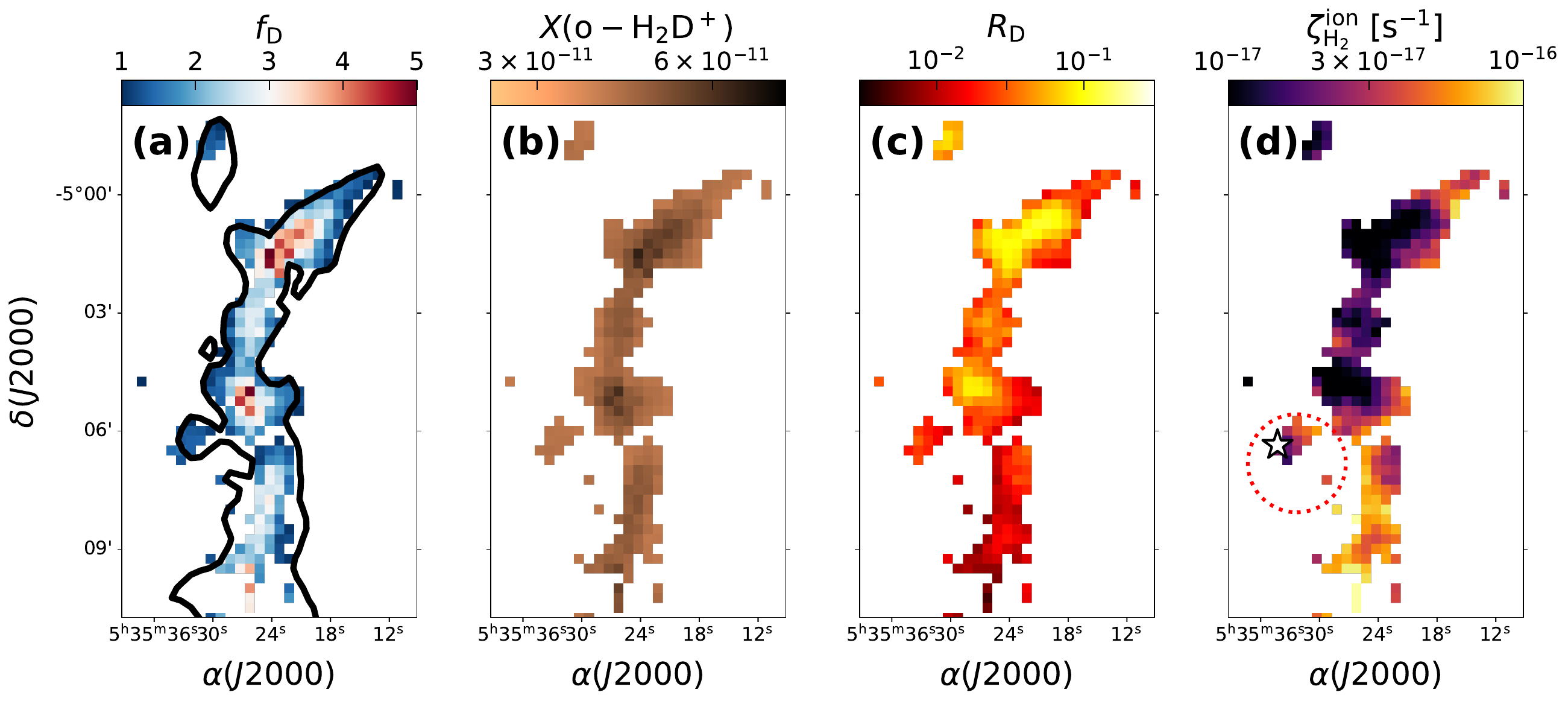}
\caption{Results of the present study for OMC-2 and OMC-3. \textbf{Panel (a)}: depletion map ($f_\mathrm{D}$) obtained from the ratio between the expected abundance of C$^{18}$O ($X_\mathrm{0}(\mathrm{C^{18}O})$) in Orion and the measured one ($N_\mathrm{tot}(\mathrm{C^{18}O})/N(\mathrm{H_2})$). The black contours represent $I(\mathrm{N_2H^+})>4$~K~km~s$^{-1}$ (signal-to-noise of 10, see Fig.~\ref{fig:aux_maps}, panel b). \textbf{Panel (b)}: abundance of o$-$H$_2$D$^+$ ($X(\mathrm{o-H_2D^+})$) inferred from the C$^{18}$O depletion factor using the correlation from \citet{saba20}. \textbf{Panel (c)}: deuteration fraction, determined as $R_\mathrm{D} = N(\mathrm{DCO^+})/N(\mathrm{HCO^+})$, across OMC-2 and OMC-3 (see text for its determination). \textbf{Panel (d)}: $\crir$ determined with Eq.~(\ref{eq:saba_eq}). The white star represents a B-type star within the selected region boundaries. The red dotted circle is a candidate expanding CO-shell \citep{fedde18} possibly associated to the star.}
\label{fig:data}
\end{figure*}

The ratio between the expected abundance of C$^{18}$O at the galactocentric distance of Orion \citep[i.e., $X_0(\mathrm{C^{18}O})=1.8\times10^{-7}$;][]{gianne17} and the one measured ($X_\mathrm{obs} = N_\mathrm{tot}(\mathrm{C^{18}O})/N(\mathrm{H_2})$) gives the depletion factor:
\begin{equation}
    f_{\rm{D}} = \frac{X_0}{X_{\rm{obs}}}.
\end{equation}
Figure~\ref{fig:data}, panel (a), shows this depletion factor, which, despite our source selection (see Sect.~\ref{subsec:sourcesel}), is extended for $\sim1$~pc from OMC-2 to the tip of OMC-3. Depletion factors $f_\mathrm{D}>1$ correlate extremely well with N$_2$H$^+$ intensities above $I(\mathrm{N_2H^+})>4$~K~km~s$^{-1}$ (signal-to-noise of 10, black contours) and with the column densities $N(\mathrm{H_2})\gtrsim10^{22}$~cm$^{-2}$ across the two regions (see Fig.~\ref{fig:aux_maps}). Our map shows depletion factors ranging within $f_\mathrm{D}\sim1-5$, consistently with previous studies at comparable resolution (i.e., $\sim30''$) in infrared dark clouds \citep[IRDCs; e.g.,][]{herna11,saba19}, single low- \citep[e.g.,][]{crapsi05} and high-mass \citep[e.g.,][]{fontani06} star-forming regions. Similarly to these studies, we also determine $f_\mathrm{D}$ slightly below 1 ($\sim0.85$ on average), which would imply an over-abundance of C$^{18}$O compared to the expected $X_0$. Since these fields are restricted to the edges of the map and given the small deviations from unity, such $f_\mathrm{D}<1$ could result from uncertainties in their derivation rather than a physical over-abundance of $\mathrm{C^{18}O}$.
We have therefore masked their corresponding fields in Fig.~\ref{fig:data}, panel (a), to consider only the regions where depletion is effective ($f_\mathrm{D}\geq1$), which constrains our map to the filamentary structure extended throughout OMC-2 and OMC-3.

\subsection{Expected abundance of o$-$H$_2$D$^+$ on parsec scales}\label{subsec:h2dabund}

The challenging observations of o$-$H$_2$D$^+$ have limited its study to surveys of individual cores and clumps within low- \citep[e.g.][]{caselli08,bovino21} and high-mass \citep[e.g.][]{saba20,reda21b, reda22} star-forming regions. The opportunity to proxy its abundance with $f_\mathrm{D}$, under proper conditions, opens up the path for parsec-scale studies of the ionisation processes. We thus apply the empirical correlation between $f_\mathrm{D}$ and $X(\mathrm{o-H_2D^+})$ determined by \citet{saba20}:
\begin{equation}
    {\rm{log}}_{\rm{10}}(X({\rm{o-H_2D^+}})) = 0.05\times f_{\rm{D}} - 10.46
    \label{eq:sabatini20}
\end{equation}
Figure~\ref{fig:data}, panel (b) shows the $X(\mathrm{o-H_2D^+})$ estimated with Eq.~(\ref{eq:sabatini20}). The dynamic range of $X(\mathrm{o-H_2D^+})$ is constrained within a factor of 3 ($\sim3-8\times10^{-11}$) and in excellent agreement with the same abundances from \citet{saba20}, as expected since the depletion factors are also in close agreement. These abundances of o$-$H$_2$D$^+$ are broadly consistent with independent estimates for low-mass cores \citep{caselli08}, however, they are an order of magnitude, on average, lower than the values determined for high-mass regions \citep{reda21b,saba23}. The difference between our estimates and those in high-mass regions can be explained based on two major factors. First, the higher column densities probed in these regions ($N(\mathrm{H_2})\gtrsim5\times10^{23}$ cm$^{-2}$) compared to our work. Second, the beam dilution effect when comparing our low-resolution IRAM-30m observations ($30''$) to their high-resolution ALMA observations ($\sim2''$). A comparison between observations and 3D modelling of the CO freeze-out, in fact, suggests depletion factors of $f_\mathrm{D}\sim3-8$ to be typical at scales of $\sim0.1$~pc \citep{bovino19}. These scales are close to the resolution of our maps ($\sim0.06-0.07$~pc), thus, since $X(\mathrm{o-H_2D^+})$ is inferred from $f_\mathrm{D}$, it suffers from the same beam dilution effect. 

The $\mathrm{o-H_2D^+}$ abundances we derive are mostly in agreement with those determined in previous studies, highlighting the CO depletion factor as a viable proxy for $X(\mathrm{o-H_2D^+})$. However, independent measurements of these two parameters are needed to ultimately break the degeneracy and evaluate the empirical correlation found by \citet{saba20}. The increasing sensitivity and spatial resolution of modern astronomical facilities, e.g., ALMA, will allow us to achieve this goal in the near future.

\subsection{Deuteration fraction across OMC-2 and OMC-3}\label{subsec:deutfrac}

The deuteration fraction ($R_\mathrm{D}$) is directly connected to all D-bearing isotopologues of H$_3^+$. These molecules are in fact the main deuterium donors in the cold and dense gas of molecular clouds \citep[e.g.,][]{cecca14}. In the present study, we estimate the deuteration fraction as $N(\mathrm{DCO^+})/N(\mathrm{HCO^+})$ through the DCO$^+$ (3$-$2) and HCO$^+$ (1$-$0) lines. Our first approach could be again estimating the column densities of both tracers using Eq.~(\ref{eq:coldens}) in the optically thin limit ($\tau\ll1$) and assuming LTE at the kinetic temperature $T_\mathrm{K}$. While reasonable for C$^{18}$O, assuming LTE for DCO$^+$ and HCO$^+$ is not ideal given the larger critical densities of these two molecules \citep[e.g.,][]{sanhueza12,keown16}. In addition, the HCO$^+$ (1$-$0) can easily become optically thick \citep[e.g.,][]{vasyu12}, therefore a correction for $\tau$ is expected.

Similarly to C$^{18}$O, we inspect the temperature-corrected intensities of both HCO$^+$ and DCO$^+$, and compare them to $N(\mathrm{H_2})$ (see Fig.~\ref{fig:tracers_comparison}). $I_\mathrm{T}(\mathrm{HCO}^+)$ (panel c) shows a clear dependence with $N(\mathrm{H_2})$ with a compact distribution. Only a handful of fields show a significant scatter, suggesting self-absorbed profiles to be almost absent in our sample \citep[see][]{tafalla23}. However, while not self-absorbed, the HCO$^+$ intensities likely suffer from opacity as the distribution flattens towards $N(\mathrm{H_2})\sim10^{23}$~cm$^{-2}$. $I_\mathrm{T}(\mathrm{DCO}^+)$ (panel d) also shows a clear dependence with $N(\mathrm{H_2})$. Compared to those of HCO$^+$, the intensities of DCO$^+$ increase almost linearly with $N(\mathrm{H_2})$, for a distribution resembling the one of N$_2$H$^+$. However, $I_\mathrm{T}(\mathrm{DCO}^+)$ shows a higher degree of scatter compared to $I_\mathrm{T}(\mathrm{N_2H^+})$, suggesting deviations from the LTE assumption used to apply Eq.~(\ref{eq:tcorrint}).

To estimate the excitation effects on both DCO$^+$ and HCO$^+$, we run a grid of radiative transfer models per each pixel using RADEX \citep{vandertak07} in two different steps\footnote{The radiative transfer calculations have not been applied to C$^{18}$O because, without proper modelling of the cloud which includes a freeze-out threshold, depletion may be mistaken for opacity \citep[e.g., see][]{tafalla21}}. In the first step, we input $N_\mathrm{tot}(\mathrm{DCO^+})$ and $N_\mathrm{tot}(\mathrm{HCO^+})$ calculated in the LTE, optically thin limit and vary $n$, the volume density, in logarithmic intervals between $10^4$ and $10^6$~cm$^{-3}$. In the second step, we fix these same volume densities and let the two column densities vary by an order of magnitude around the LTE, optically thin values. We then take from the output of RADEX the two $N_\mathrm{tot}$ which minimise the observed intensities. In both steps, we input $T_\mathrm{K}$ as the kinetic temperature and fix the linewidths of DCO$^+$ and HCO$^+$ to $\Delta v=0.75\pm0.10$~km~s$^{-1}$ and $\Delta v=2.5\pm0.4$~km~s$^{-1}$, respectively. These linewidths are the mean values obtained by fitting a few representative spectra in our fields and their error is taken as the standard deviation seen across these fits.

The deuteration fraction (Fig.~\ref{fig:data}, panel c) shows more than an order of magnitude variation ($\sim0.005-0.1$) across OMC-2 and OMC-3. Such values are consistent with previous studies in high-mass star-forming regions \citep{saba20,saba23,pazu23}, using the same molecular species, and theoretical predictions from chemical models \citep{albert13}. The connection found between $f_\mathrm{D}$ and $R_\mathrm{D}$ in these and other studies \citep[e.g., in low-mass regions;][]{crapsi05} demonstrates how the deuteration process is enhanced in the cold and dense gas where CO is depleted \citep[e.g.,][for a review]{caselli12}. Simulations of turbulent and magnetised pc-scale filaments, such as the ISF \citep[e.g.,][]{pattle17,hacar18,ziel22}, show increasing levels of deuteration as the column density increases \citep{koertgen18}. These authors determine deuteration fractions within $\sim0.01-0.1$ in $t\lesssim0.4$~Myr, a timescale consistent with the expected survival time of the dense gas in Orion \citep[$\sim0.5$~Myr;][]{hacar24}. Since o$-$H$_2$D$^+$ is the driver of deuteration in the early phases of the cloud evolution \citep{bovino20}, and given the dependence between $f_\mathrm{D}$ and $R_\mathrm{D}$, a correlation also between $X(\mathrm{o-H_2D^+})$ and $f_\mathrm{D}$ would be readily explained.

\begin{figure*}
\centering
\includegraphics[width=0.9\linewidth]{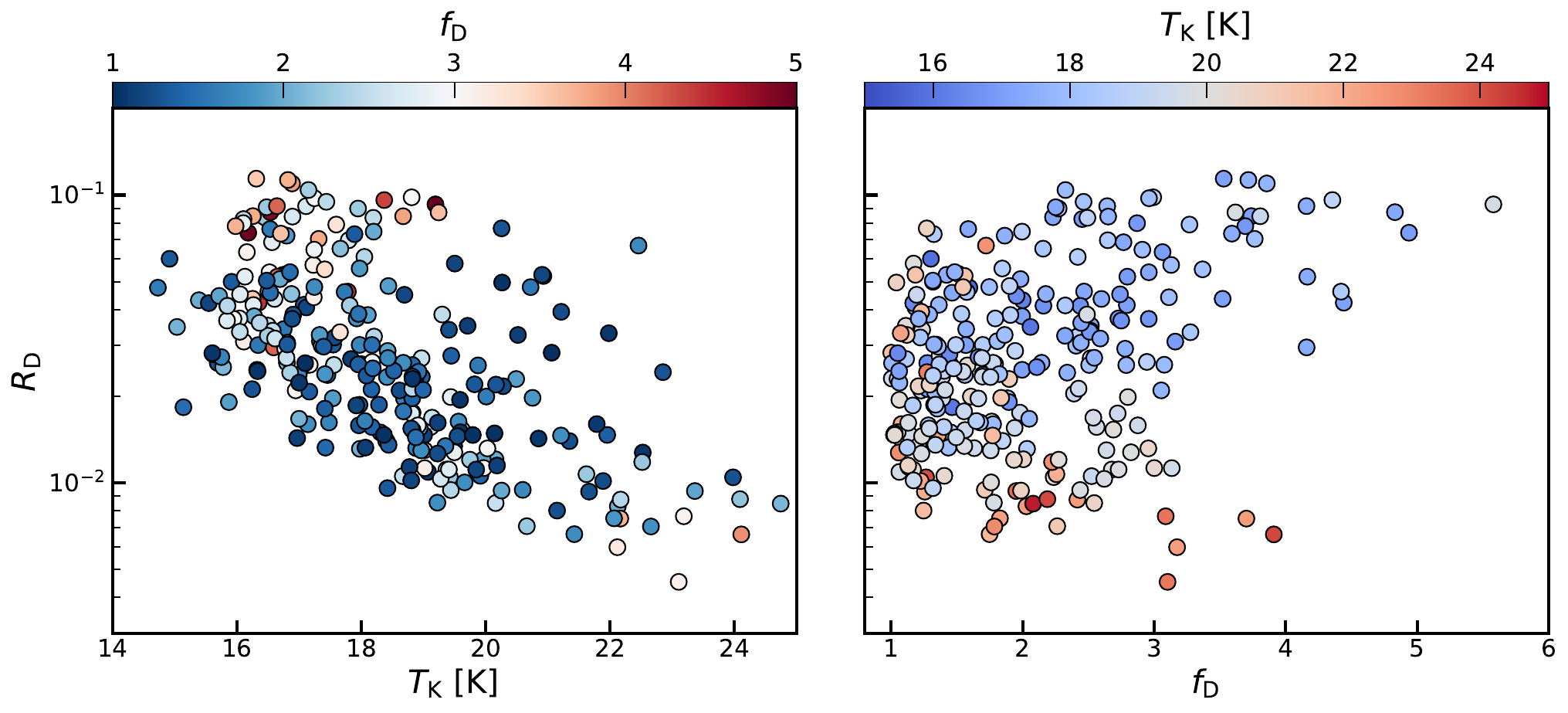}
\caption{\textbf{left panel}: deuteration fraction ($R_\mathrm{D}$) compared to the kinetic temperature ($T_\mathrm{K}$). Each point is colour-coded with the corresponding depletion factor ($f_\mathrm{D}$). \textbf{right panel}: deuteration fraction ($R_\mathrm{D}$) now compared to the depletion factor ($f_\mathrm{D}$) and colour-coded with the corresponding kinetic temperature ($T_\mathrm{K}$).}
\label{fig:tempdepen}
\end{figure*}

The variation in $R_\mathrm{D}$ across our maps shows a progressive increase from OMC-2, in the South, to OMC-3 towards the North. Such gradient reflects the morphology of the HCO$^+$ and DCO$^+$ integrated intensities seen in Fig.~\ref{fig:raw_data} (a,c). The variation could be linked to a change in volume density causing the higher-J levels of the two molecules to be more populated. Yet, the non-LTE analysis returns a limited range in volume densities, varying within $\sim0.3-2\times10^5$~cm$^{-3}$ (with mean value and standard deviation of $n=1.0\pm0.2\times10^5$~cm$^{-3}$). Therefore, the significant variation in $R_\mathrm{D}$ is likely the result of a more efficient deuteration process towards OMC-3. 

It is well known \citep[since][]{watson74} that deuteration is affected by temperature variations, which cause a change in the reaction rates. The deuteration of HCO$^+$ into DCO$^+$ is particularly sensitive to this temperature effect \citep[see][for a recent example]{pazu23}. Thus, the gradient we observe in $R_\mathrm{D}$ is possibly linked to the slight change in $T_\mathrm{K}$ from OMC-3 towards OMC-2 (see Fig.~\ref{fig:aux_maps}, panel c). We explore this hypothesis in Fig.~\ref{fig:tempdepen}, where we display the comparison of $R_\mathrm{D}$ with $T_\mathrm{K}$ (left panel) and $f_\mathrm{d}$ (right panel). The left panel shows an anti-correlation between the deuteration fraction and the kinetic temperature, already seen in previous studies \citep[e.g.,][]{pazu23}, most prominent for fields with low depletion factors ($f_\mathrm{D}\lesssim2$). When the temperatures increase towards and above $T_\mathrm{K}\gtrsim20$~K, the deuteration fraction drops to values of $\lesssim0.01$. For temperatures $\sim16-19$~K however, there is a significant scatter, for which the fields with higher depletion ($f_\mathrm{D}\gtrsim2-3$) also show a higher degree of deuteration ($R_\mathrm{D}\gtrsim0.03$). These same points can be seen in the right panel as they show a positive correlation between $R_\mathrm{D}$ and $f_\mathrm{D}$. This dependence between the two parameters has been previously reported by \citet{crapsi05} for a similar range of values and volume density regime ($n\sim10^5$~cm$^{-3}$). We therefore expect the correlation between $R_\mathrm{D}$ and $f_\mathrm{D}$ to be robust below $\lesssim20$~K, and become more loose for higher temperatures making the corresponding $\crir$ estimates possibly uncertain.

As a final test, we determine a scaling relation between $R_\mathrm{D}$ and $T_\mathrm{K}$ in Fig.~\ref{fig:fitandcompar}. First, we sample the median value, and Inter-Quartile Range (IQR), of $R_\mathrm{D}$ and $T_\mathrm{K}$ in bins of one~Kelvin (red circles and error bars, respectively). Then, we fit a power law dependence to these estimates, for which the best fit (black line) reads as $R_\mathrm{D}\sim T_\mathrm{K}^{-4}$ (see Eq.~\ref{eq:RdTk}). Such sharp decrease of $R_\mathrm{D}(\mathrm{HCO}^+)$ with $T_\mathrm{K}$ has been previously observed across different clouds \citep[e.g.,][]{pazu23}, although never before condensed in a scaling relation. The direct comparison of this scaling relation with estimates of $R_\mathrm{D}(\mathrm{HCO^+})$ and $T_\mathrm{K}$ in massive clumps \citep{mietti11} and high-mass regions \citep{gerner15} confirms the goodness of our fit (see full discussion in Appendix~\ref{sec:deutfit}). The systematic nature of the effect, which appears as independent from the star-formation regime of the region, suggests the use of temperature-corrected deuteration fractions for future estimates of $\crir$ using the method described in Sect.~\ref{sec:method}.

\subsection{The destruction rate $k^\mathrm{o-H_3^+}_\mathrm{CO}$}\label{subsec:desrate}

Equation~\ref{eq:saba_eq} assumes o-H$_3^+$ to be primarily destroyed by CO \citep{bovino20, reda24}. While this is one of the main processes in dense cores, other destruction pathways can occur when studying the ionisation rate at parsec scales. In particular, electrons can reduce the H$_3^+$ abundance through dissociative recombination. This process becomes important both at extreme densities, when the CO is almost entirely depleted \citep[e.g.,][]{reda24}, and at lower densities, especially in a region externally illuminated by stellar radiation, such as the ISF \citep[e.g.,][]{pabst19}.

We thus consider estimates of the ionisation fraction, $x(\mathrm{e}^-)$, in these two density regimes and compare them with the analytical limit $f_\mathrm{D} < 9\times10^{-7}/x(\mathrm{e^-})$ proposed by \citet{reda24} for Eq.~(\ref{eq:saba_eq}). First, we consider the estimates in massive dense cores, including some in Orion \citep{bergin99}. These authors estimate ionisation fractions $x(\mathrm{e^-})\lesssim10^{-7}$ at $N(\mathrm{H_2})\gtrsim10^{22}$~cm$^{-2}$ for these sources, leading to a depletion limit for the method of $f_\mathrm{D}\gtrsim9$. Second, we consider the estimates in OMC-2 and OMC-3 using the C$_2$H and HCN lines \citep{salas21}. These authors determine ionisation fractions $x(\mathrm{e}^-)\leq3\times10^{-6}$ at $n\sim5\times10^3$~cm$^{-3}$, leading to a non-physical $f_\mathrm{D}>0.3$. Given our range in both column and volume density ($N(\mathrm{H_2})\sim10^{22}-10^{23}$~cm$^{-2}$ and $n(\mathrm{H_2})\sim0.3-2\times10^5$~cm$^{-3}$, respectively), we can safely assume the destruction through CO as main process affecting the H$_3^+$ abundance for our degree of depletion.

We can finally determine the destruction rate $k^\mathrm{o-H_3^+}_\mathrm{CO}$ in our regions. We computed the latter pixel-by-pixel using $T_\mathrm{K}$ and the coefficients found in the KInetic Database for Astrochemistry \citep[KIDA\footnote{\textbf{\url{https://kida.astrochem-tools.org}}};][]{wake12} for the ion-polar reaction \mbox{$\mathrm{H_3^+} + \mathrm{CO}\rightarrow\mathrm{HCO^+} + \mathrm{H_2}$}. The adopted $k^\mathrm{o-H_3^+}_\mathrm{CO}$ rates, in the temperature range of 10-25~K, are relatively constant and lie within $\sim2.15-2.35\times10^{-9}$~cm$^{-3}$~s$^{-1}$.

\subsection{Error budget in the determination of $\crir$}\label{subsec:errorbud}

Our indirect method to determine $\crir$ presents a number of sources of error, which will be discussed in the following. Aside from $X(\mathrm{o-H_2D^+})$, for which we assume the CO depletion is an efficient proxy (see Sect.~\ref{sec:method}), all the other parameters in Eq.~(\ref{eq:saba_eq}) bear a corresponding uncertainty.
\begin{itemize}
    \item $k_\mathrm{CO}^\mathrm{o-H_3^+}$, $N(\mathrm{C^{18}O})$: both only depend on the temperature (see Eq.~\ref{eq:c18odens} and Sect.~\ref{subsec:desrate}), thus we propagated their error accordingly using the uncertainty on $T_\mathrm{K}$ \citep{friesen17};
    \item $[^{16}\mathrm{O}]/[^{18}\mathrm{O}]$: used to convert $N(\mathrm{C^{18}O})$ to $N(\mathrm{CO})$ in Eq.~(\ref{eq:saba_eq}), it is equal to $560\pm26$ for the Local ISM \citep[see Table 4 in][]{wilson94};
    \item $R_\mathrm{D}$: given its definition and its derivation with RADEX, we consider as total uncertainty the combined uncertainties on the DCO$^+$ and HCO$^+$ linewidths (see Sect.~\ref{subsec:deutfrac});
    \item $l$: we assume the typical size of N$_2$H$^+$ structures in OMC-2 and OMC-3 (see Sect.~\ref{sec:method}), along with its corresponding error (see also Socci et al. submitted).
\end{itemize}

The error per field in our analysis reaches up to $\sim50\%$ of the corresponding $\crir$ across OMC-2 and OMC-3. However, this error contains a major flat contribution coming from the conservative choices made to estimate the uncertainty on some parameters (e.g., linewidths of DCO$^+$ and HCO$^+$). A more thorough analysis would therefore significantly reduce the uncertainty in several of our fields. Since the determination of $\crir$ remains indirect and this error budget does not hinder our further discussion and conclusions, we stick to these choices for the present analysis.

\subsection{The Orion cosmic-ray ionisation rate map}\label{subsec:zeta}

The map of $\crir$ (Fig.~\ref{fig:data}, panel d) shows a dynamic range of more than one order of magnitude across the OMC-2/OMC-3 region. The global South-North decrease in $\crir$ is likely connected to the gradient seen in the deuteration fraction, therefore to the effect of temperature (see Fig.~\ref{fig:tempdepen}, left panel). However, the local variations are likely induced by the change in $f_{\rm D}$ along the filament (see Fig.~\ref{fig:tempdepen}, right panel). As a result of both effects, $\crir$ varies within $\sim5\times10^{-18}-2\times10^{-16}$~s$^{-1}$ within the two regions.

The median value we derive of $\crir\sim2.4\times10^{-17}$~s$^{-1}$ is in good agreement with the estimates in high-mass star-forming regions using the same method \citep{saba20,saba23}. The slight differences reflect the variation in $X(\mathrm{o-H_2D^+})$ (see Sect.~\ref{subsec:h2dabund}), thus we can ascribe these differences to a resolution effect. Among the $\crir$ we determine, its highest values (i.e., $\gtrsim2\times10^{-16}$~s$^{-1}$) are seen towards the South of the map and, in particular, in the region close to the OMC-2 FIR4 protocluster \citep[e.g.,][]{johnstone03} and the B-type star HD~37060 (see Fig.~\ref{fig:data}, panel d). The presence of these two heating sources can indirectly affect our estimate through an increase in $T_\mathrm{K}$, with a corresponding decrease of $R_\mathrm{D}$. In addition, the outflows from the protocluster \citep[e.g.,][]{lopez13,Chahine22} and the UV-irradiation from the star could promote the photodissociation of C$^{18}$O, inducing an increase in $f_\mathrm{D}$ (see Fig.~\ref{fig:tempdepen}, right panel). However, higher values of $\crir$ could also result from a local enhancement in the CR flux \citep[see][]{padovani15,padovani16,padovani19} caused by these objects as suggested by previous findings: values of CRs ionisation rates as high as $\crir\sim10^{-14}$~s$^{-1}$ has been previously reported in OMC-2 FIR4 \citep{cecca14b,fontani17}; a candidate expanding CO-shell has been detected in close correspondence of the B-type star \citep{fedde18}, for which CRs could travel beyond its edge and promote the ionisation in the South of our filament.

Despite all the influences to the ionisation rate, $\crir\approx10^{-17}$~s$^{-1}$, originally found by \citet{spi68}, has been instead considered for several decades as the reference value for the ISM. While such value is reproduced by our results, our pc-scale map shows at least an order of magnitude variation for $\crir$ across OMC-2 and OMC-3. No constant value for the ionisation rate is therefore expected, but instead, local changes in the physical properties of the gas play a crucial role in its variation. Among the possible physical influences on the CR ionisation rate, we consider here the effect of increasing $N(\mathrm{H_2})$ as the prime driver of its variation.

\section{Discussion}\label{sec:discussion}

\begin{figure*}
\centering
\includegraphics[width=0.7\linewidth]{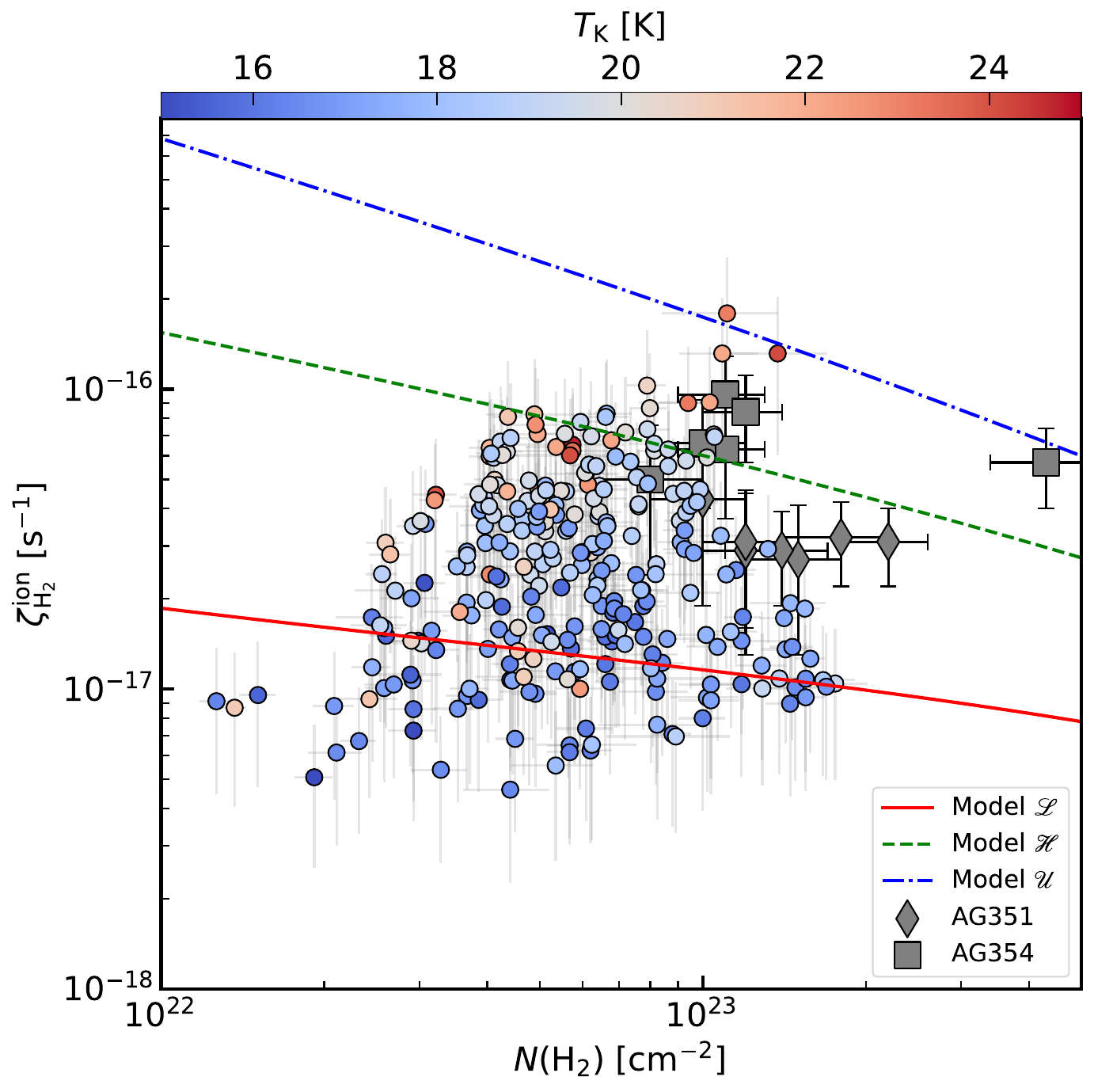}
\caption{Comparison between $\crir$ and the column density $N(\mathrm{H_2})$. The theoretical predictions for the CR propagation in molecular clouds are displayed as colour-coded lines \citep[blue, green, and red lines;][]{pado22}, while the $\crir$ estimates in high-mass regions are displayed in grey \citep[squares and diamonds;][]{saba23}. The estimates of $\crir$ from our study are displayed as circles, colour-coded by $T_\mathrm{K}$.}
\label{fig:corrected_zeta}
\end{figure*}

The different models for energy loss of CR particles (protons, electrons and heavy nuclei) with increasing $N(\mathrm{H_2})$ have been originally studied by \citet{pado09}. These models account for the ionisation by primary CR protons and electrons, plus secondary electrons from primary CRs. For the column density regime we probe, the models of CR propagation (see Fig.~\ref{fig:corrected_zeta}, coloured lines) are largely dominated by CR protons and their secondaries. As a general trend, all models predict a decrease of $\crir$ with increasing $N(\mathrm{H_2})$. Different slopes come from the comparison between the CR propagation models and observational results: model $\mathscr{L}$, from the CR spectrum observed by the Voyager missions \citep{cummings16,stone19}; model $\mathscr{H}$, from the average ionisation rate estimated in diffuse clouds \citep{indi12,neufeld17}; model $\mathscr{U}$, from the upper limit of the ionisation rate in diffuse clouds. Observational results show a decreasing trend as well, however, the large number of methods and tracers used to infer $\crir$ produces a significant scatter in the data, especially for dense clouds \citep[see Fig.~B.1 in][]{padovani24}.

Our new results considerably increase the statistics for the available $\crir$ estimates compared to previous studies \citep[e.g.,][]{caselli98,vandertak00}, probing a column density regime barely explored before. Figure \ref{fig:corrected_zeta}, shows the comparison between $\crir$ and $N(\mathrm{H_2})$, colour-coded by $T_\mathrm{K}$, with their corresponding errors \citep[see][and Sect.~\ref{subsec:errorbud}, respectively]{lombardi14}. 
In addition, we plot the models from \citet{pado22} for different slopes of the CR proton energy spectrum (coloured lines). Almost the entire sample of $\crir$ determined in our study is enclosed within the model $\mathscr{L}$ and the model $\mathscr{H}$, with only a fraction of fields outside of this range. Despite their departure from the bulk of our points, these ionisation rates above the model $\mathscr{H}$ and below the model $\mathscr{L}$ are still consistent with the theoretical predictions within the errors. None of the predicted propagation models clearly reproduces our results. Our estimates show, however, the functional dependence between $\crir$ and $N(\mathrm{H_2})$ expected from these models. While the temperature dependence of $R_\mathrm{D}$ (see Sect.~\ref{subsec:deutfrac}) contributes to the scatter, the fields with $T_\mathrm{K}\lesssim20$~K show ionisation rates decreasing with $N(\mathrm{H_2})$.

As we include the $\crir$ determined in two high-mass star-forming regions using the same method \citep[diamonds and squares;][]{saba23}, the ionisation rates across the two studies are in close agreement. Overall our estimates show values closer to those in AG531, while only the highest ionisation rates we determine are comparable to the ones in AG354. The combined distribution of ionisation rates from the two studies is continuous with $N(\mathrm{H_2})$ and show a mild functional dependence across column densities $\sim10^{22}-5\times10^{23}$~cm$^{-2}$. The method proposed by \citet{bovino20} therefore allows for homogeneous estimates of $\crir$ in a large range of scales and column density regimes, when o$-$H$_2$D$^+$ is either successfully detected or reasonably inferred.

The two estimates from the present paper and \citet{saba23} are in close agreement, however spurious effects have to be considered in their determination. While resolution on the one hand affects the degree of depletion (see Sect.~\ref{subsec:h2dabund}), on the other hand it also influences the minimum physical size we can probe with our observations. Such physical size is condensed in the path length $l$ from Eq.~(\ref{eq:saba_eq}). As mentioned in Sect.~\ref{sec:method}, measuring either the line-of-sight length or the volume density of the tracers is complex, therefore $l$ is usually taken equal to the radial size of the structures identified in o$-$H$_2$D$^+$ \citep{saba20,saba23}. We explore the influence of $l$ on the $\crir$ by measuring the typical radial size of the DCO$^+$ emission, our least extended tracer. We therefore perform a radial sampling on the DCO$^+$ integrated intensity map following the peaks of its emission and fit the average profile with a Gaussian function (see Sect.~\ref{sec:addmaps} for a full discussion). The FWHM of the profile ($0.18\pm0.15$~pc) is taken as representative $l$ to compute the estimate of $\crir$. Being $3-4$ times the $l$ adopted in our analysis (0.05~pc; see Sect.~\ref{sec:method}), this FWHM results in $\crir$ values on average lower by the same factor (see Eq.~\ref{eq:saba_eq}, $\crir\propto 1/l$). The median ionisation rate with $l\sim0.18$~pc is $\crir\sim7\times10^{-18}$~s$^{-1}$ and the majority of the fields is below the model $\mathscr{L}$. Ionisation rates as low as $\sim10^{-18}$~s$^{-1}$, produced by correspondingly low-energy CRs, are unrealistic when considering the degree of feedback present in the ISF \citep{pabst19} and strongly disagree with independent estimates, which indeed suggest higher values \citep[e.g.,][]{fontani17,salas21}. As $f_\mathrm{D}$ and $R_\mathrm{D}$ are positively correlated, their ratio is only mildly affected by resolution effects. The path length, $l$, on the other hand, suffers directly from a poor resolution, thus representing the main limitation in the determination of ionisation rates with Eq.~(\ref{eq:saba_eq}) \citep[see also][for different ways to estimate $l$ and a detailed analysis on its effect to the $\crir$]{bovino20,reda24}.

Our results highlight the CO depletion as available proxy for the o$-$H$_2$D$^+$ abundance at parsec scales. However, caution should be taken when interpreting these results: first, determining $f_\mathrm{D}$ can be tricky and the use of multiple CO lines is usually advised \citep[e.g.,][]{saba22}; second, the connection between $f_\mathrm{D}$ and the o$-$H$_2$D$^+$ abundance has been derived at sub-parsec scales \citep{saba20} and works within a constrained range of $T_\mathrm{K}$, outside which multiple processes may become temperature dependent (e.g., degree of deuteration). In turn, o$-$H$_2$D$^+$-based estimates are confirmed as an effective method to homogeneously sample $\crir$ across more than an order of magnitude in column density. Ultimately, in the absence of local sources of CRs, $N(\mathrm{H_2})$ alone appears to drive the ionisation degree in molecular clouds \citep{pado22}. Finally, our results also highlight the importance of high resolution observations when estimating the degree of depletion and to reliably estimate $l$, which can significantly affect the $\crir$ determined. Independent interferometric observations of both CO isotopologs and o$-$H$_2$D$^+$ are needed to validate the previous correlation by \citet{saba20} and provide a more robust estimate of the cosmic ray ionization rate.

\section{Conclusions}\label{sec:conclusions}

In this work, we present the first parsec-scale estimate of the cosmic-ray ionisation rate determined by employing the most recent analytical framework proposed by \citet{bovino20}. To this end, we presented novel IRAM-30m observations of the Integral Shape Filament in Orion using the following suite of tracers: C$^{18}$O~(2$-$1), DCO$^+$~(3$-$2) and HCO$^+$~(1$-$0). We first determined the C$^{18}$O depletion factor under density and temperature regimes for which we expect significant freeze-out and no photodissociation effects. The C$^{18}$O depletion factor was then used to proxy the o$-$H$_2$D$^+$ abundance, following the correlation determined by \citet{saba20}, and ultimately determine the ionisation rate $\crir$. The main results of the study are the following:
\begin{enumerate}
    \item The depletion factor map is extended from the North of OMC-2 to the tip of OMC-3 with values ranging within $f_\mathrm{D}\sim1-5$. Such dynamic range is consistent with previous estimates across whole clouds \citep{herna11,saba19,feng20} and within single star-forming regions \citep{crapsi05,fontani06,saba22,saba23}.
    \item The o$-$H$_2$D$^+$ abundance inferred from $f_\mathrm{D}$ has a constrained dynamic range between $\sim3-8\times10^{-11}$, in agreement with previous independent observations \citep{caselli08,gianne19,mietti20,reda21b,saba23}. Although estimated indirectly, the good agreement of $X(\mathrm{o-H_2D^+})$ across studies suggests $f_{\rm D}$ to be a reliable proxy, at least at the studied spatial scales.
    \item The deuteration fraction ($R_\mathrm{D}$) we determine shows values within $\sim0.005-0.1$. These values are consistent with previous estimates using the same molecular species \citep{saba23} and show a positive correlation with $f_\mathrm{D}$ for the volume density regime we explore ($n\sim10^5$~cm$^{-3}$), consistently with previous works \citep{crapsi05}. The deuteration fraction shows a temperature dependence, most prominent for $T_\mathrm{K}\gtrsim20$~K. Below this value, its positive correlation with $f_\mathrm{D}$ is more robust. 
    \item The map of $\crir$ shows a South-North gradient with values varying within $\sim5\times10^{-18}-2\times10^{-16}$~s$^{-1}$. Such gradient is likely connected to the temperature dependence of $R_\mathrm{D}$, however, local variations, especially for $T_\mathrm{K}\sim16-19$~K are driven by the degree of depletion. Overall the large dynamic range in $\crir$ suggests local physical properties (e.g., $N(\mathrm{H_2})$) to influence its value, as expected from theoretical models \citep{pado09}.
    \item The ionisation rates determined for the column density regime probed ($\sim10^{22}-2\times10^{23}$~cm$^{-2}$) are comparable to the theoretical predictions for the same column densities \citep{pado22}. They show a mild anti-correlation with the column density, as expected from the models of CRs propagation within clouds, extended to values of $N(\mathrm{H_2})\gtrsim10^{23}$~cm$^{-2}$ when complemented by the estimates of \citet{saba23}. 
    \item Finally, we tested the influence of the estimated path length $l$ on $\crir$. When directly determined from our low-resolution DCO$^+$ observations, the path length $l$ is $\sim0.18$~pc. Such value, when applied to our analysis, produces results no longer consistent with those in high-mass regions \citep{saba23} and with independent studies in the ISF \citep[e.g.,][]{fontani17}. In addition, these lower ionisation rates are only marginally reproduced by the theoretical predictions \citep{pado22}. All these discrepancies suggest the need of high resolution data, to provide reliable estimates of $l$, which represents the major limiting factor in the ionisation rates determination.
\end{enumerate}

\begin{acknowledgements}
This project has received funding from the European Research Council (ERC) under the European Union’s Horizon 2020 research and innovation programme (Grant agreement No. 851435).
GS acknowledges the projects PRIN-MUR 2020 MUR BEYOND-2p (``Astrochemistry beyond the second period elements'', Prot. 2020AFB3FX) and INAF-Minigrant 2023 TRIESTE (``TRacing the chemIcal hEritage of our originS: from proTostars to planEts''; PI: G. Sabatini). SB acknowledges the ANID BASAL project FB210003.
This work is based on IRAM-30m telescope observations carried out under project numbers 032-13. IRAM is supported by INSU/CNRS (France), MPG (Germany), and IGN (Spain). 
This research has made use of the SIMBAD database, operated at CDS, Strasbourg, France.
This research has made use of NASA’s Astrophysics Data System.
\end{acknowledgements}

\nocite{*}
\bibliographystyle{aa}
\bibliography{bibl.bib}

\begin{appendix}

\section{Additional Maps}\label{sec:addmaps}

\begin{figure*}
\centering
\includegraphics[width=\linewidth]{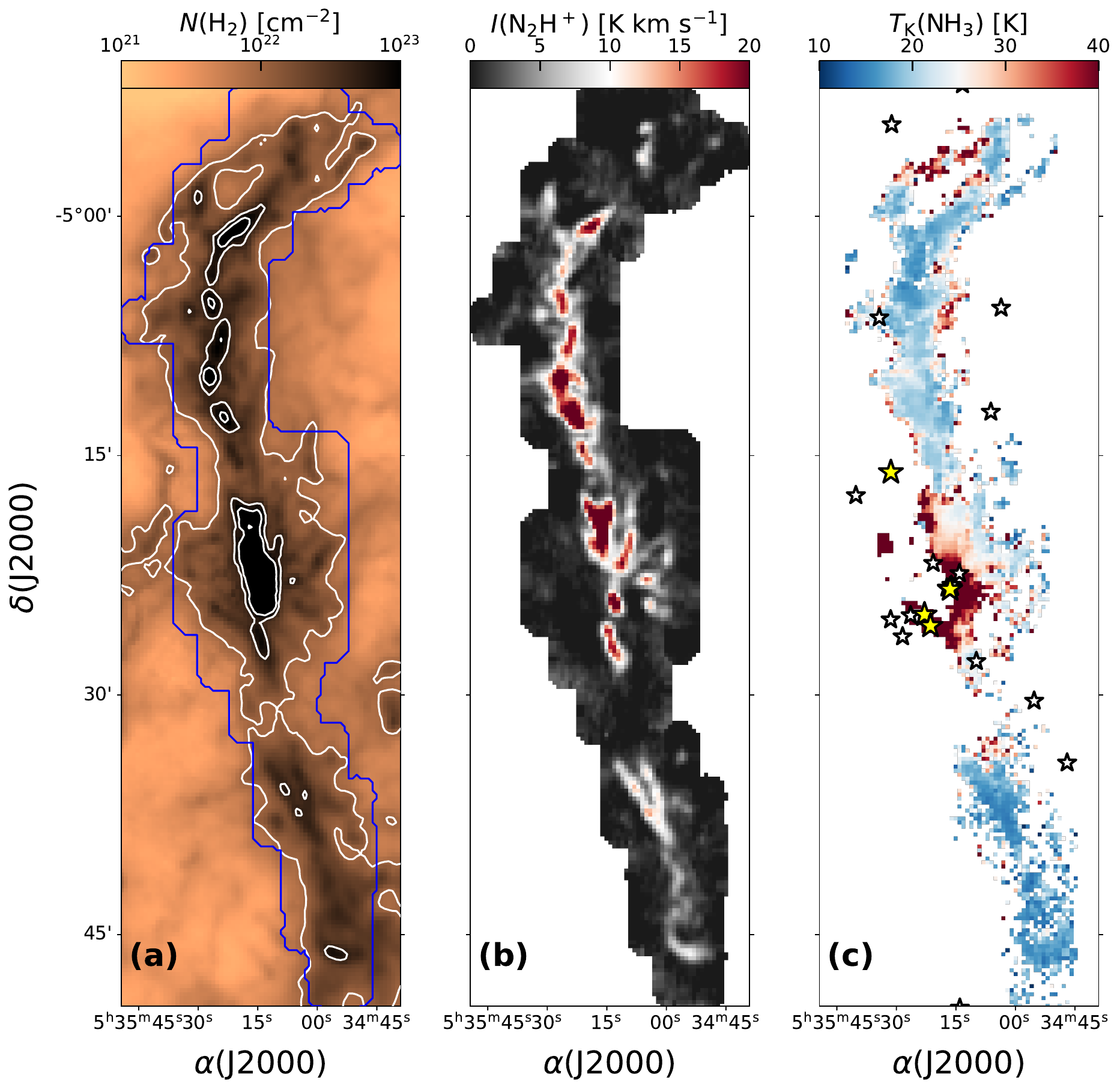}
\caption{The ISF as seen in different types of emission. \textbf{Panel (a)}: column density map of H$_2$ ($N(\mathrm{H_2})$) obtained from dust opacity \textit{Herschel}+\textit{Planck} observations at 850 $\mu$m \citep{lombardi14}. The white contours correspond to [$10^{22}$, $5\times10^{22}$, $10^{23}$]~cm$^{-2}$, while the blue contour represents the footprint of our IRAM-30m observations. \textbf{Panel (b)}: N$_2$H$^+$ (1$-$0) integrated intensity map of observed with IRAM-30m at 30$''$ resolution \citep[see][for additional information]{hacar17}. \textbf{Panel (c)}: kinetic temperature of the gas ($T_\mathrm{K}$) obtained from observations of the two inversion transitions NH$_3$ (1,1) and NH$_3$ (2,2) at 32$''$ with the GBT \citep{friesen17}. Here, the white and yellow stars represent the B-type and O-type stars in the region, respectively, same as Fig.~\ref{fig:raw_data}, panel (b).}
\label{fig:aux_maps}
\end{figure*}

\begin{figure*}
\centering
\includegraphics[width=\linewidth]{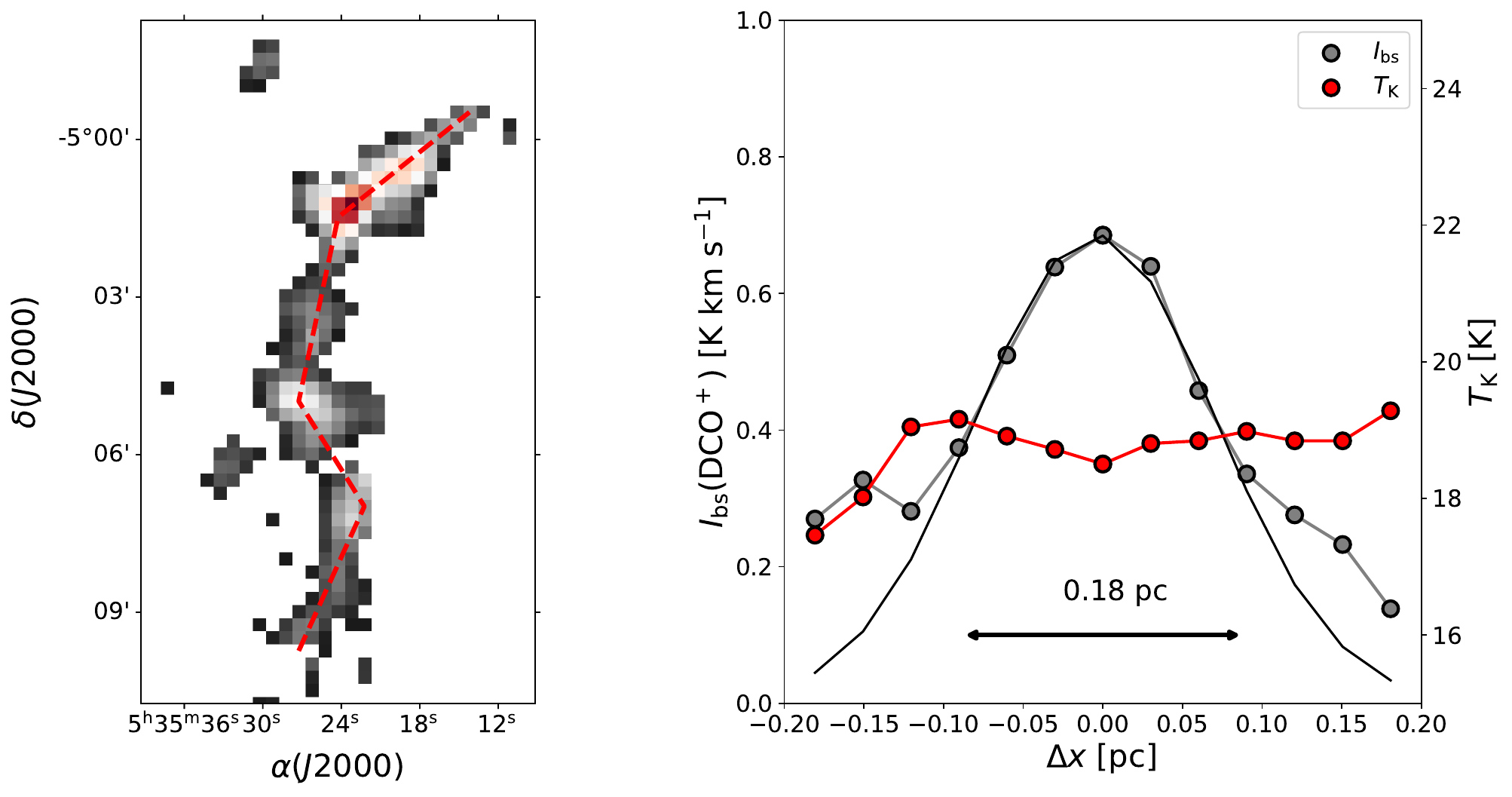}
    \caption{\textbf{Left panel}: integrated intensity map of DCO$^+$ (3$-$2) (see also Fig.~\ref{fig:raw_data}), masked based on the criteria discussed in Sects.~\ref{subsec:sourcesel},\ref{subsec:depletion}. On top of the map, the axis used for the radial sampling is displayed (red dashed line). \textbf{Right panel}: average intensity ($I_\mathrm{bs})$ and temperature ($T_\mathrm{K}$) radial profiles computed across the red line in the left panel. The Gaussian best fit to the profile is also displayed (solid black line), along with the derived FWHM (solid arrow).}
\label{fig:radialsampling}
\end{figure*}

Figure \ref{fig:aux_maps} shows the OMC-1/-2/-3/-4 regions as seen through the total column density of H$_2$ \citep[panel a;][]{lombardi14}, through the N$_2$H$^+$ (1$-$0) integrated emission \citep[panel b;][]{hacar17} and through the kinetic temperature \citep[panel c;][]{friesen17}. The latter was directly delivered by the authors alongside with the corresponding error\footnote{\textbf{\url{https://greenbankobservatory.org/science/gbt-surveys/gas-survey-2/}}}, while the other two have been computed as follows:
\begin{itemize}
    \item $N(\mathrm{H_2})$: first, we determine the visual extinction in the K-band ($A_\mathrm{K}$) from the dust opacity at 850~$\mu$m ($\tau_\mathrm{850}$):
    \begin{equation}\label{eq:akcal}
        A_{\rm{K}} = \gamma\times\tau_{\rm{850}} + \delta,
    \end{equation}
    where $\gamma = 2460$~mag and $\delta = 0.012$~mag for Orion A\footnote{The linear correlation between the two parameters holds across the majority of Orion A, aside for the region close to the Trapezium stars. Since we have excluded OMC-1 from the analysis, we can safely apply Eq.~(\ref{eq:akcal}) for our study \citep[see][for a full discussion]{lombardi14}.}. Then we convert the extinction in the K-band to visual extinction ($A_\mathrm{V}$) using $A_\mathrm{K}/A_\mathrm{V} = 0.112$ \citep{rieke85}. Finally, we obtain $N(\mathrm{H_2})$ using the conversion \citep{bohlin78}:
    \begin{equation}
        N({\rm{H_2}}) = 0.94\times10^{21}\times A_{\rm{V}}~{\rm{cm}}^{-2}.
    \end{equation}
    The corresponding error on the column density of H$_2$ per pixel is computed following the same procedure and using the error on $\tau_\mathrm{850}$.
    \item $I(\mathrm{N_2H^+})$: the intensity map of N$_2$H$^+$ (1$-$0) is obtained by integrating all the hyperfine components of the line \citep[e.g.,][]{caselli95} in the velocity range $\Delta V_\mathrm{lsr} = [-5, 20]$~km~s$^{-1}$.
\end{itemize}

Figure~\ref{fig:radialsampling} shows the radial sampling results onto the DCO$^+$ (3$-$2) integrated intensity map to determine the path length $l$ (see Sect.~\ref{sec:discussion}). To perform the radial profile fitting, we employed the following approach: first, we mask the integrated intensity map using the same criteria applied in Sects.~\ref{subsec:sourcesel},\ref{subsec:depletion}; second, we subtract a baseline of $3\times \sigma(\mathrm{DCO^+})$ to the profile (obtaining $I_\mathrm{bs}$); third, we draw the axis manually onto the map following the peaks of emission; fourth, we sample all 4 segments with perpendicular cuts spaced by $\theta_\mathrm{beam}$ and we sample the map along the cuts each pixel (i.e., each $\theta_\mathrm{beam}$); finally, we average the 4 profiles and fit the resulting radial profile with a Gaussian function. The average profile width is $\mathrm{FWHM}=0.18\pm0.15$~pc.

\section{The deuteration fraction scaling relation}\label{sec:deutfit}

\begin{figure}
\centering
\includegraphics[width=\linewidth]{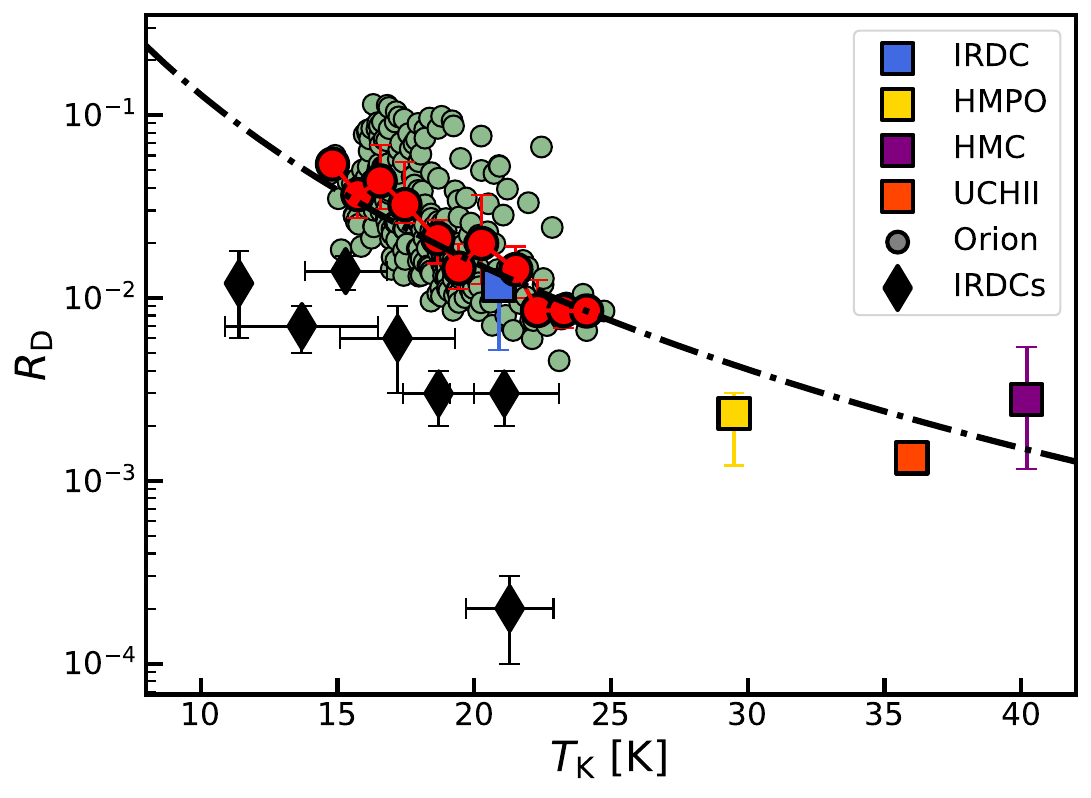}
\caption{Estimates of $R_\mathrm{D} = N(\mathrm{HCO^+})/N(\mathrm{DCO^+})$ and $T_\mathrm{K}$ across different studies: estimates in Orion, both single fields and median per bin (green and red circles, respectively; this study), in IRDCs \citep[black diamonds;][]{mietti11}, in high-mass star-forming regions \citep[colour-coded squares;][see text for discussion on $T_\mathrm{K}$]{gerner14,gerner15}. The black line represents the best fit to the median values of our estimates in Orion and it is exemplified by Eq.~\ref{eq:RdTk}.}
\label{fig:fitandcompar}
\end{figure}

Figure~\ref{fig:fitandcompar} shows the correlation between $R_\mathrm{D}$ and $T_\mathrm{K}$ in our study and in comparison with previous estimates. To reduce the scatter of our fields, we have sampled the median value, with corresponding IQR, of both $R_\mathrm{D}$ and $T_\mathrm{K}$ in bins of one~Kelvin within $14-25$~K (red circles and error bars, respectively). Then, we fitted these 11 median values with a power law dependence of the form $R_\mathrm{D} = R_\mathrm{D}^0\times(T_\mathrm{K}/10~\mathrm{K})^{a}$, where 10~K has been chosen as arbitrary normalisation temperature. The best fit to this dependency reads as follows (black line):
\begin{equation}\label{eq:RdTk}
    R_{\rm{D}}=0.24\times\bigg(\frac{T_{\rm{K}}}{10~{\rm{K}}}\bigg)^{-3.8},
\end{equation}
suggesting variations of two orders of magnitude already within $\sim10-40$~K, temperature range seen across the ISF (see Fig.~\ref{fig:aux_maps}, panel c).

To further asses the reliability of this scaling relation, we directly compare the scaling relation determined with deuteration fractions from previous studies. To perform an homogeneous comparison, we have selected deuteration fractions determined as $N(\mathrm{DCO^+})/N(\mathrm{HCO^+})$ and with an independent measurement of $T_\mathrm{K}$. Thus, the first comparison is against the estimates of $R_\mathrm{D}$ and $T_\mathrm{K}$ in massive clumps embedded within IR Dark Clouds \citep[IRDCs, black diamonds in Fig.~\ref{fig:fitandcompar};][Tables~5-9]{mietti11}. The deuteration fractions in these clumps show a similar scaling with the temperature as our estimates in Orion, up to $\sim22$~K. While the scaling is similar, the absolute values are, on average, a factor of $\sim5$ lower, which suggests an enhancement of HCO$^+$ deuteration in our fields. Given the similar resolution in the observations ($\sim20''-30''$), column density ($\sim10^{22}-10^{23}$~cm$^{-2}$) and volume density ($\sim0.3-2\times10^5$~cm$^{-3}$) across the two studies, these higher deuteration fractions are possibly connected to, overall, warmer temperatures in our fields, plus the possible presence of unresolved depletion and density gradients within the massive clumps \citep{mietti11}.

We now explore whether or not the deuteration fraction variation is connected to the star-formation regime of the region. Thus, the second comparison is against the estimates of $R_\mathrm{D}$ and $T_\mathrm{K}$ in high-mass regions in different evolutionary stages \citep[i.e., High-Mass Protostellar Objects (HMPO), Cores (HMC) and Ultra-Compact HII regions (UCHII), colour-coded squares in Fig.~\ref{fig:fitandcompar};][]{gerner15}. We have gathered the column densities of DCO$^+$ and HCO$^+$ from CDS\footnote{\textbf{\url{https://cdsarc.cds.unistra.fr/viz-bin/cat/J/A+A/579/A80}}}, computed the median and IQR per evolutionary stage, and associated to them their average kinetic temperature determined from the chemical models of \citet{gerner14}\footnote{\citet{gerner15} also provides updated kinetic temperatures for a chemical network including deuterated species. However, their models do not improve significantly the results of \citet{gerner14} and still do not converge for some of the molecular species in the chemical network. Thus, we opted for the $T_\mathrm{K}$ from \citet{gerner14} given the more general set of reactions considered.}. Across the different evolutionary stages, the high-mass regions show an excellent agreement with the scaling relation determined in Eq.~\ref{eq:RdTk}. The prime factor driving the decrease of $R_\mathrm{D}(\mathrm{HCO^+})$ in molecular clouds appears therefore to be $T_\mathrm{K}$, while other factors, such as the star-formation history of the regions, only marginally affect its variation.

\end{appendix}

\end{document}